\batchmode

\documentclass[jgrga]{agutex}

\usepackage{graphicx}
\usepackage{color}
\usepackage{url}

\authorrunninghead{R. MANZANAS ET AL.}
\titlerunninghead{ASSESSING EL NI\~NO-DRIVEN SKILL}
\authoraddr{Corresponding author: R. Manzanas \\ Grupo de Meteorolog\'ia. Instituto de F\'isica de Cantabria. CSIC-Univ. de Cantabria, Avda. de los Castros, s/n, 39005, Santander, Spain \\ Email: rmanzanas@ifca.unican.es}

\begin{document}

\title{GLOBAL FORTY-YEARS VALIDATION OF SEASONAL PRECIPITATION FORECASTS: ASSESSING EL NI\~NO-DRIVEN SKILL}

\authors{R. MANZANAS\altaffilmark{1}, M.D. FR\'IAS\altaffilmark{2}, A.S. COFI\~NO\altaffilmark{2} and J.M. GUTI\'ERREZ\altaffilmark{1}}

\altaffiltext{1}{Grupo de Meteorolog\'ia. Instituto de F\'isica de Cantabria. CSIC-Univ. de Cantabria, Avda. de los Castros, s/n, 39005, Santander, Spain.}
\altaffiltext{2}{Grupo de Meteorolog\'ia. Dpto. Matem\'atica Aplicada y Ciencias de la Computaci\'on. Univ. de Cantabria, Avda. de los Castros, s/n, 39005, Santander, Spain.}

\begin{abstract}
The skill of seasonal precipitation forecasts is assessed worldwide ---grid point by grid point--- for the forty-years period 1961-2000. To this aim, the ENSEMBLES multi-model hindcast is considered. Although predictability varies with region, season and lead-time, results indicate that 1) significant skill is mainly located in the tropics ---$20$ to $40$\% of the total land areas---, 2) overall, SON (MAM) is the most (less) skillful season and 3) predictability does not decrease noticeably from one to four months lead-time ---this is so especially in northern south America and the Malay archipelago, which seem to be the most skillful regions of the world---.
An analysis of teleconnections revealed that most of the skillful zones exhibit significant teleconnections with El Ni\~{n}o. Furthermore, models are shown to reproduce similar teleconnection patterns to those observed, especially in SON ---with spatial correlations of around 0.6 in the tropics---. Moreover, these correlations are systematically higher for the skillful areas.
Our results indicate that the skill found might be determined to a great extent by the models' ability to properly reproduce the observed El Ni\~{n}o teleconnections, i.e.,  the better a model simulates the El Ni\~{n}o teleconnections, the higher its performance is. 
\end{abstract}

%
%

%

\begin{article}
%
%

\section{Introduction}\label{s.intro}
Seasonal forecasting is a promising research field with enormous impact on different socio-economic sectors  such as water resources, agriculture, energy and health  \citep[see][and references therein]{Doblas-Reyes_2013}. 
Nowadays, seasonal forecasts are routinely produced by several institutions around the world using different global ocean-atmosphere coupled models. Moreover, these products are collected by a number of regional focal points worldwide to produce operational consensus seasonal forecasts with socio-economic potential ---see, e.g., the Regional Climate Outlook Forum (RCOF) sponsored by the World Meteorological Organization (WMO), \url{http://www.wmo.int/pages/prog/wcp/wcasp/clips/outlooks/climate_forecasts.html}--- .  
However, there are still several limiting factors 
which hinder the practical use of seasonal forecasts 
\citep[see, e.g.,][]{Goddard_2010}. 
For instance, it is known that seasonal predictability strongly varies with the target variable, region and season \citep[see, e.g.,][]{Halpert1992,vanOldenborgh2004,Barnston2010,DoblasReyes2010}.

Therefore, in order to properly communicate the uncertainties related to seasonal predictions, it is needed to develop a comprehensive assessment of the performance of the different forecasting models worldwide, especially for those variables most widely used by the stakeholders and end-users.
In particular, precipitation is the most challenging case for being less skillfully predicted than surface temperatures \citep[see, e.g.,][]{Barnston2010,DoblasReyes2010,Bundel2011}.
However, the majority of verification studies for seasonal forecasts of this variable have been conducted over limited areas of the world and for concrete seasons \citep[see, e.g.,][]{Batte2011,Lim2011,Kim2012a,Landman2012}.
A few studies have also been conducted worldwide \citep[][]{vanOldenborgh2005,Wang2009,Barnston2010,DoblasReyes2010}, using a number of validation scores ---correlation, Ranked Probability Skill Score (RPSS) and Brier Skill Scores (BSS)---. However, the limited hindcast period available in the latter works does not ensure a robust statistical validation. For instance, \citet{DoblasReyes2010} analyzed the ENSEMBLES multi-model seasonal dataset, computing averaged scores over six large-scale regions of the world for the period 1991-2005.

In this paper we present a global ---grid point by grid point--- forty-years (1961--2000) robust validation of the ENSEMBLES multi-model seasonal hindcast ---the longest-to-date available dataset of retrospective forecasts--- by applying a simple tercile-based probabilistic validation scheme, obtaining a simple and easy to interpret ---adequate for communication with decision-makers--- measure of skill, the ROC Skill Score, which is the only 
 skill score recommended by the Lead Centre for the Standardized Verification System (SVS) of Long Range Forecasts (LRF) \url{http://www.bom.gov.au/wmo/lrfvs/index.html} for the verification of probabilistic seasonal forecasts.
One- and four-months lead predictions are considered for each of the four standard boreal seasons.
Besides, since ENSO is known to be the major driving factor for seasonal predictability \citep{Goddard_Dilley_2005}, 
we also analyze the El Ni\~no-driven component of the skill in the different regions by assessing the ability of the models to properly reproduce the observed teleconnections.
Thus, the two main goals of this study are 1) to fill the lack of an up-to-date user-oriented global validation of seasonal precipitation forecasts considering a long (forty years) period ---identifying those regions of the world with significant skill--- and 2) to analyze the El Ni\~no-driven component of this skill, assessing to which extent predictability might be determined by this phenomenon. 

The paper is organized as follows: The data used are described in Sec. \ref{s.data}. The methodology applied is explained in Sec. \ref{s.methodology}.
Results are discussed through Sec. \ref{s.results}.
Finally, the main conclusions are given in Sec. \ref{s.conclusions}.

\section{Data}\label{s.data}
\begin{figure*}[htb]\begin{center}
\noindent\includegraphics[width=0.7 \linewidth]{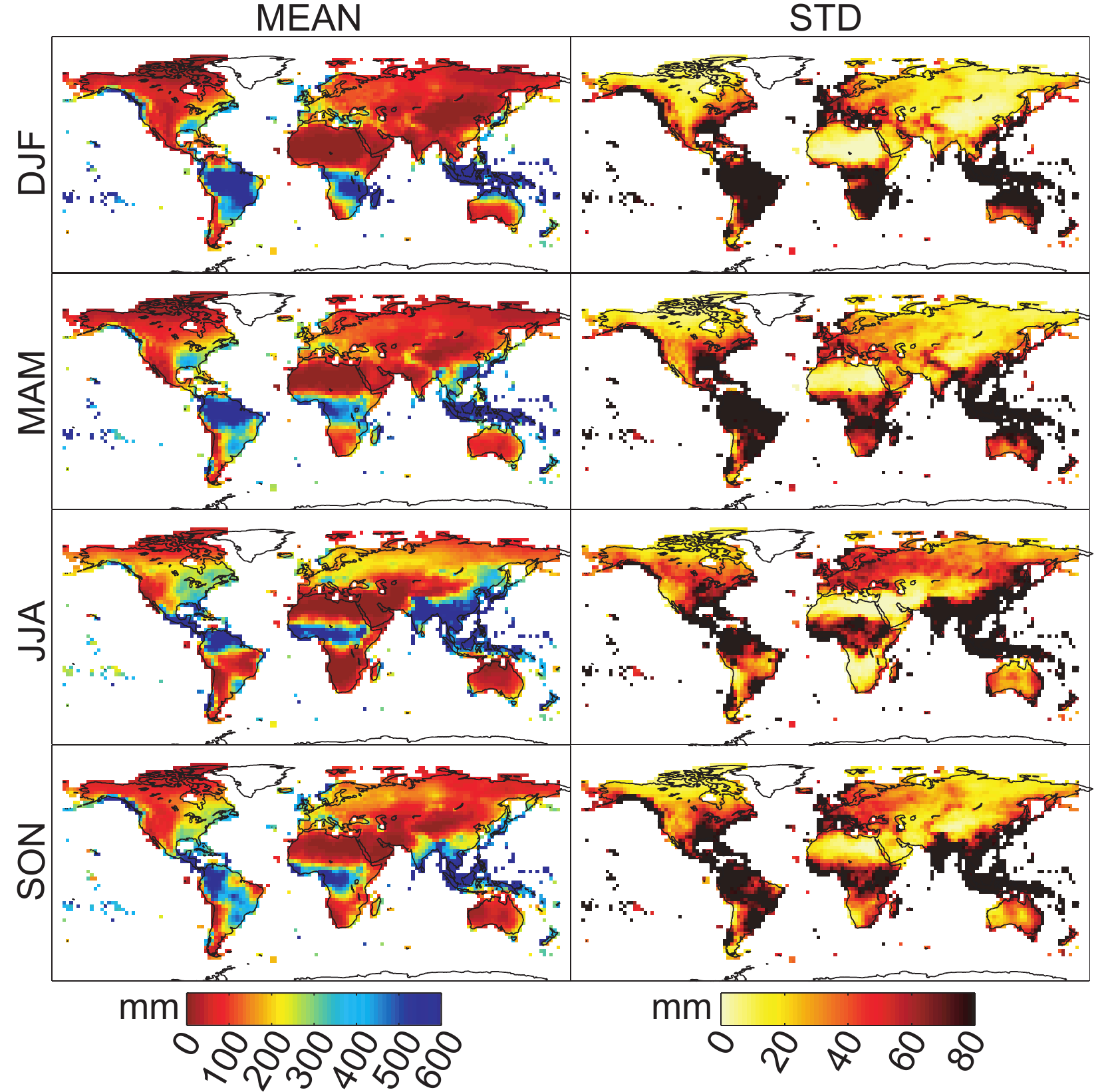}
\caption{Mean and inter-annual standard deviation (left and right columns, respectively) of seasonal accumulated precipitation from VASClimO v1.1 for the four seasons considered (in rows) in the period 1961-2000.}
	\label{f.obs}\end{center}
\end{figure*}

On the one hand, VASClimO v1.1 \citep{Beck2005} was considered as the reference dataset for validation. This gauge-based product provides monthly precipitation totals on a  $2.5^\circ$ resolution grid for the global land areas (except the Antarctica) for the period 1951-2000. Fig. \ref{f.obs} shows the mean seasonal totals and the corresponding inter-annual standard deviation (STD) for this dataset for the period of study 1961-2000.

In order to test the sensitivity to the reference data in the validation process, all calculations were also done for an alternative precipitation dataset, the Global Precipitation Climatology Centre full data reanalysis version 6 (GPCC v6) \citep{Becker2013}. The results obtained in both cases were very similar, thus only VASClimO v1.1 is considered hereafter.

On the other hand, predictions were obtained from the longest-to-date multi-model seasonal hindcast, provided by the EU project ENSEMBLES \citep{Weisheimer2009}, which comprises five state-of-the-art atmosphere-ocean coupled models from the following centres: The UK Met Office (UKMO), M\'{e}t\'{e}o France (MF), the European Centre for Medium-Range Weather Forecasts (ECMWF), the Leibniz Institute of Marine Sciences (IFM-GEOMAR) and the Euro-Mediterranean Centre for Climate Change (CMCC-INGV).
Table \ref{t.models} summarizes the main components of these models.

\begin{table*}[htb]
\caption{Main components of the five state-of-the-art atmosphere-ocean coupled models contributing to the ENSEMBLES multi-model seasonal hindcast.}
\label{t.models}
\begin{tabular}{c c c}
\hline
{\bf Centre}	&   {\bf Atmospheric model and resolution}	&	{\bf Ocean model and resolution}	\\
\hline
{ECMWF}	&	    IFS CY31R1 (T159/L62)	&	    HOPE ($0.3^{\circ}-1.4^{\circ}$/L29)	   	\\
{IFM-GEOMAR}	&	    ECHAM5 (T63/L31)	&	    MPI-OM1 ($1.5^{\circ}$/L40)   	\\
{CMCC-INGV}	&	    ECHAM5 (T63/L19)	&	    OPA8.2 ($2.0^{\circ}$/L31)  	    	\\
{MF}	&	    ARPEGE4.6 (T63)	&	    OPA8.2 ($2.0^{\circ}$/L31)      	\\
{UKMO}	&	    HadGEM2-A (N96/L38)	&	    HadGEM2-O ($0.33^{\circ}-1.0^{\circ}$/L20)     	\\
\hline
\end{tabular}
\end{table*}

The atmosphere and the ocean were initialized using realistic estimates of their observed states and each model was run from an ensemble of nine initial conditions (nine equiprobable members). 
For each model, seven months-long runs were issued four times a year within the period 1960-2005, starting the first of February, May, August and November \citep[see][for more details about the experiment]{Weisheimer2009}. Thus, the seasons considered for validation were the standard boreal winter (DJF), spring (MAM), summer (JJA) and autumn (SON), since this allows to analyze one- and four-months lead predictions ---e.g., the initializations of August and May can be used to forecast SON---. Note that although alternative three-months seasons could be more informative in particular regions of the world, there would  be a single  lead-time available for them, thus limiting the study.
The validation period considered was 1961-2000, common to VASClimO v1.1 observations and the ENSEMBLES models. All the models were bi-linearly interpolated to the grid of the observations ---similar results were obtained using the nearest grid point interpolation technique (not shown)---.

\section{Methodology}\label{s.methodology}
The validation methodology used in this work is the tercile-based probabilistic approach previously applied in other studies \citep[see, e.g.,][]{Frias2010,Vellinga2012}. 
Thus, for each particular grid point and each particular model, member and season, the forty-years inter-annual series of predicted seasonal precipitation were categorized into three categories (dry, normal and wet), according to their respective climatological terciles within the period 1961-2000. Then, a probabilistic forecast was computed year by year by considering the number of members falling within each category, out of a total of $n=9$ members.  The terciles were defined independently for each model, considering the inter-annual series of its nine members (a total of $40 \times 9 = 360$ values) ---terciles were not computed at a member-level since no significant overlap among the dry and wet terciles of the nine members was found applying a Student's t-test---. In the case of the multi-model (denoted hereafter as MM), $n=45$ members were used to compute the  probabilistic forecasts, thus assuming equal weights for all the models. In this case, the terciles were computed independently for each model.  Note that working with precipitation categories instead of with raw values implicitly entails a bias correction grid point by grid point, which is required  for a fair validation since the different models exhibit diverse season and region-dependent biases (not shown).

Rather than using deterministic scores \citep[e.g.,][]{vanOldenborgh2005,Batte2011,Lim2011,Li2012,Singh2012}, the forecast performance is assessed in terms of a simple and easy to interpret measure of probabilistic skill \citep[see][for an introduction to verification of probabilistic forecasts]{Jolliffe2003}, the ROC Skill Score (ROCSS) \citep[see, e.g.,][]{kharin_roc_2003}. This score is a reasonable first choice to communicate the value of a forecast to the end-users \citep[see, e.g.][]{Thiaw1999}. Moreover, it is the only skill score recommended by the Lead Centre for the Standardized Verification System (SVS) of Long Range Forecasts (LRF) \url{http://www.bom.gov.au/wmo/lrfvs/index.html} for the verification of probabilistic seasonal forecasts. The ROCSS is computed as $2\,A-1$, where $A$ is the area under the ROC curve (commonly used to evaluate the performance of probabilistic systems). For each tercile category (e.g., dry events), the ROC curve is drawn as the rate of occurrences correctly forecast (the Hit Rate, $HIR_q$) versus the rate of non-occurrences incorrectly forecast (the False Alarm Rate, $FAR_q$), as a function of a varying threshold $q$ ---$q$ ranging from $0$ to $1$---. This threshold is used to convert the probabilistic prediction $p$ into a binary deterministic one ---the event is predicted to occur when $p > q$---. The ROCSS ranges from 1 (perfect forecast system) to -1 (perfectly wrong forecast system). A value zero indicates no skill with respect to a climatological prediction.  In this work, the statistical significance of the ROCSS was obtained by bootstrapping \citep{Mason2002} with 1000 samples, i.e., by generating 1000 time series of probabilistic forecasts by randomly re-sampling the original 1961-2000 sequence.

\begin{figure*}[tb]
\noindent\includegraphics[width=0.9 \linewidth]{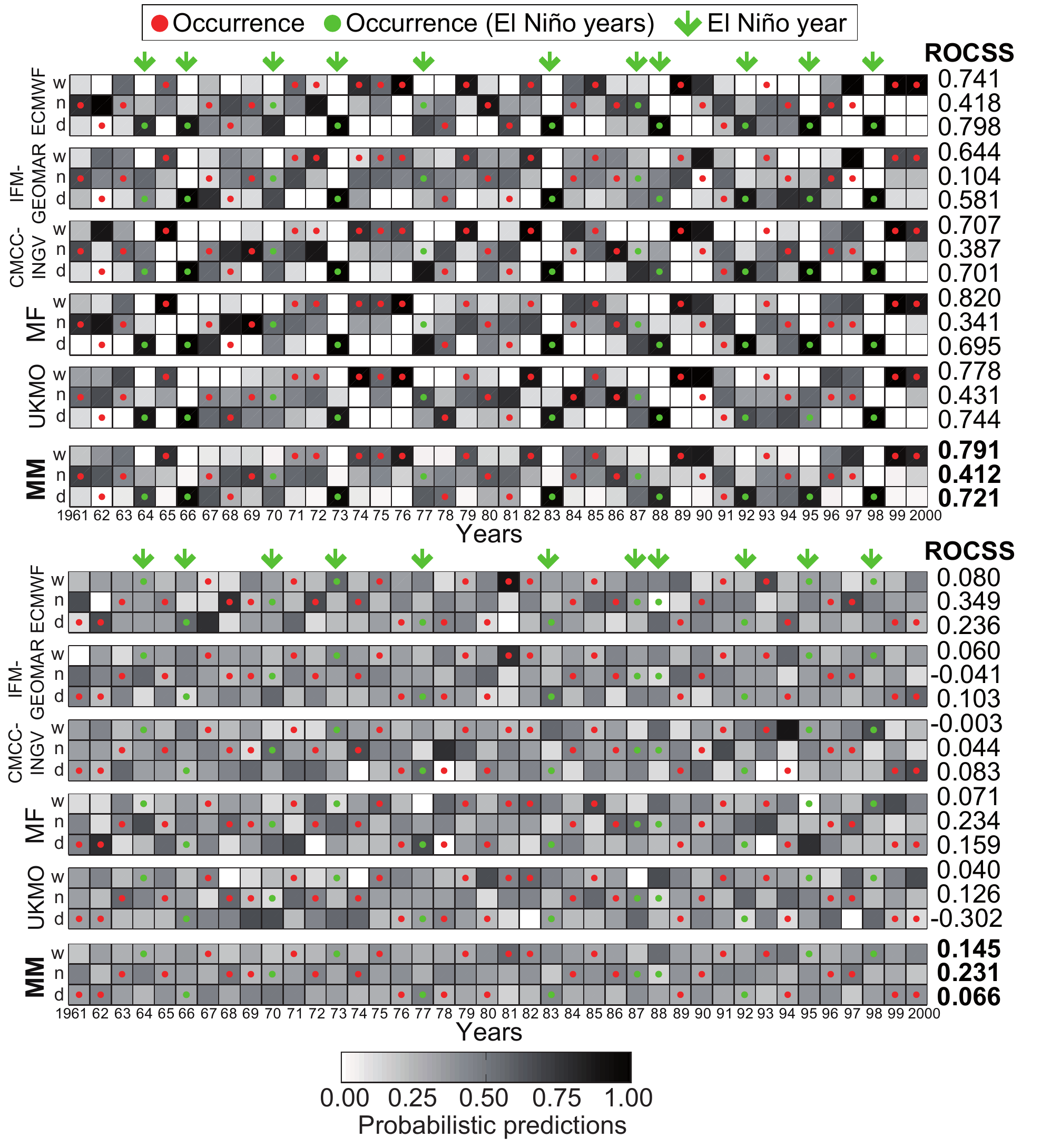}
\caption{One-month lead probabilistic predictions from the five models and the MM for SON in an illustrative grid point in (top) the Malay archipelago ---$1.25^{\circ}$\,S, $121.5^{\circ}$\,E--- and (bottom) Europe ---$51.25^{\circ}$\,N, $16.25^{\circ}$\,E---. For each tercile ---\textit{d}, \textit{n} and \textit{w} stand for dry, normal and wet, respectively---, probabilities are displayed in a white (0)-to-black (1) scale. Red (green) points mark the observed tercile in non-El Ni\~no (El Ni\~no) years ---the latter are indicated by arrows---. Numbers on the right show the ROCSS for each model and terciles.}
\label{f.cofinograma}
\end{figure*}

As an illustrative example of the validation scheme followed, Fig. \ref{f.cofinograma} shows the 1961-2000 inter-annual time-series of probabilistic predictions from the five models and the MM and the binary occurrence/non occurrence for the three terciles in two particular grid points ---one in the Malay archipelago (top panel) and the other in Europe (bottom panel)--- at one month lead-time for SON.
Although varying from year to year and from model to model, predictions exhibit a higher resolution (probabilities far from $1/3$) in the former point.
Furthermore, resolution in this case increases in general in El Ni\~{n}o years (marked with green arrows), what suggests the existence of a predictability signal linked to the latter phenomenon in this region of the world and for this season. Numbers on the right correspond to the ROCSS for the different models and terciles. High skill ---over 0.7 in most of the cases--- is found for the dry and wet terciles for the point in the Malay archipelago. On the contrary, almost no skill ---ROCSS near to zero--- is found for the point in Europe. 

\section{Results and Discussion}\label{s.results}
\subsection{Overall Skill}\label{s.global_skill}
\begin{figure*}[htb]\begin{center}
\noindent\includegraphics[width=0.95 \linewidth]{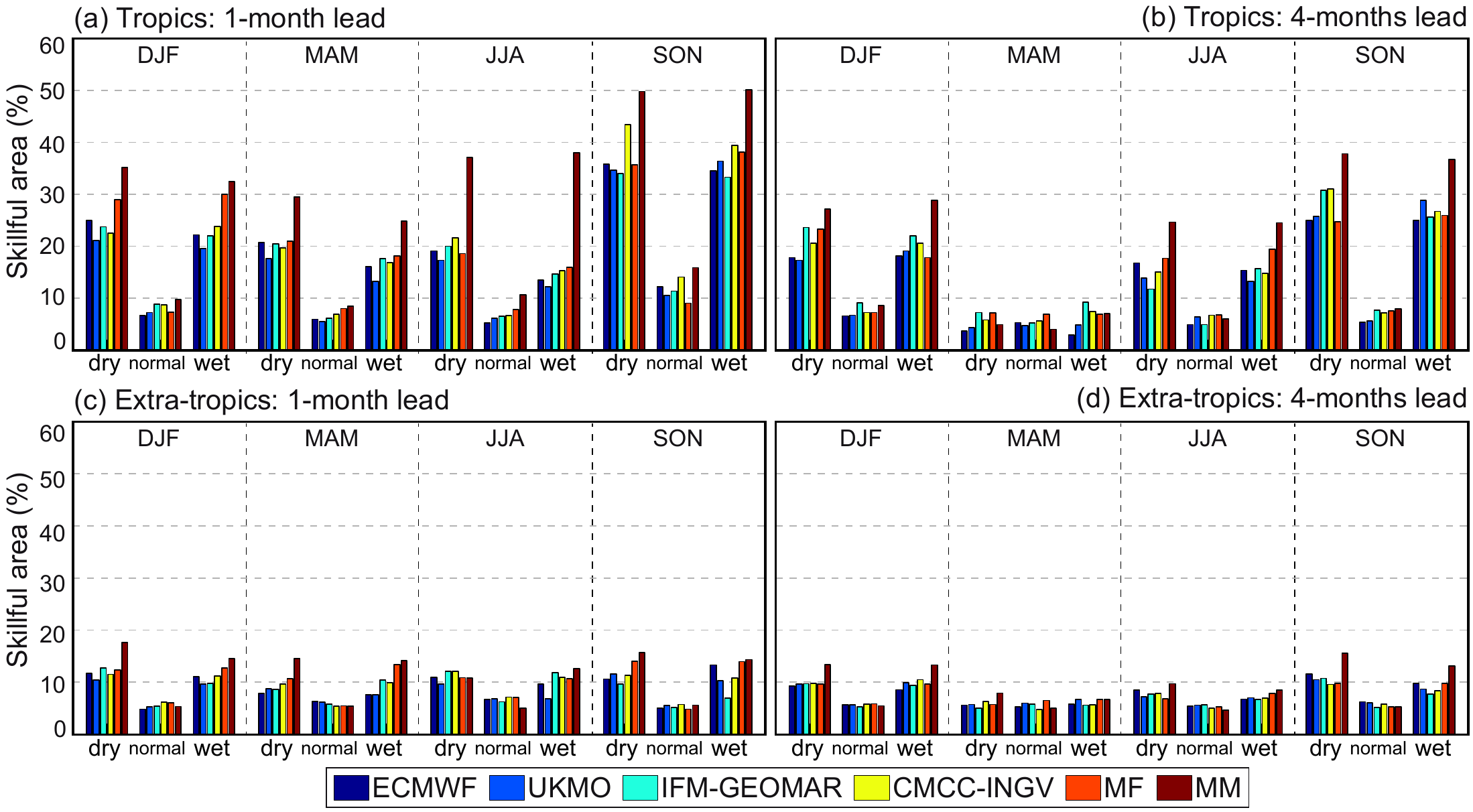}
\caption{Percentage of land areas where significant ---at a 5\% level--- ROCSS are obtained for one- and four-months lead predictions from the five models and the MM (see colors in legend) over the total in (a,b) the tropics and (c,d) the extra-tropics.}
\label{f.skill_area}\end{center}
\end{figure*}

We applied the above described methodology globally ---grid point by grid point--- in order to compute the ROCSS (and its corresponding significance) for the five models and the MM in the period 1961-2000, obtaining thus a measure of overall skill.  
As a summary of the results obtained, Fig. \ref{f.skill_area} shows the percentage of grid points with significant ---at a 0.05 level--- skill (over the total) in the land areas within the tropics ---region in between $23.5^{\circ}$\,N and $23.5^{\circ}$\,S latitudes--- and the extra-tropics, for both one- and four- months lead predictions.
Although predictability varies with region, season, model and lead-time, several general conclusions can be obtained. First, the skill concentrates in the extreme (wet or dry) terciles ---similar levels are found for both---, whereas almost no signal is obtained for normal conditions (note that the percentage of significant grid points is around 5\% in this case, which can be explained by chance according to the significance level considered). Second, predictability is mainly located in the tropics ($20$ to $40$\% of significant skillful grid points) rather than in the extra-tropics (only $10$\%), which is in agreement with previous studies \citep[see, e.g.,][]{vanOldenborgh2005}.  Furthermore, SON (MAM) is the season exhibiting the highest (lowest) percentage of skillful areas in the former region. Third, all models yield similar results for a concrete region, season and lead-time, with the MM outperforming any of the single models in all cases, which is also in agreement with previous studies \citep[see, e.g.,][]{DoblasReyes2009,Bundel2011,Ma2012}. Finally, the percentage of skillful areas at four months lead-time do not decrease noticeably with respect to the one-month lead case in any season except MAM. 
 
\begin{figure*}[htb]\begin{center}
\noindent\includegraphics[width=1.0 \linewidth]{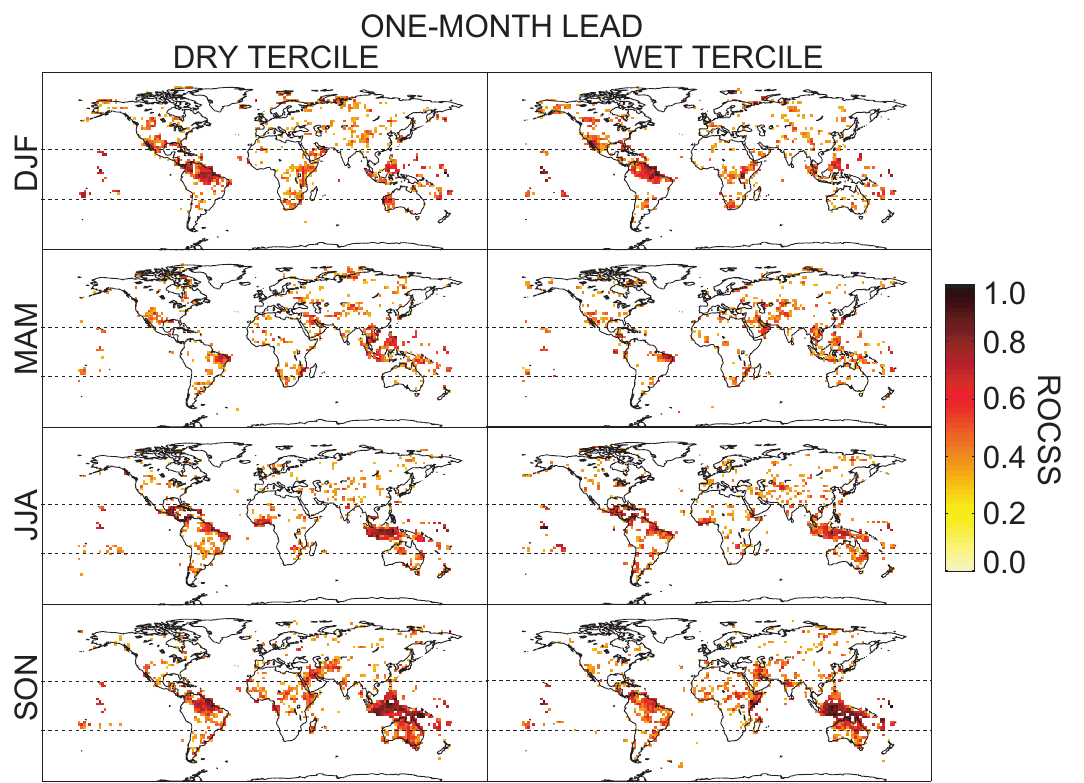}
\caption{MM skill for the dry and wet (left and right columns, respectively) terciles at one month lead-time for the period 1961-2000, by seasons (in rows).
Only significant ---at a 5\% level --- ROCSS are shown.
Dashed lines indicate the tropics/extra-tropics division.}
\label{f.skill_mm_lt1}\end{center}
\end{figure*}

\begin{figure*}[htb]\begin{center}
\noindent\includegraphics[width=1.0 \linewidth]{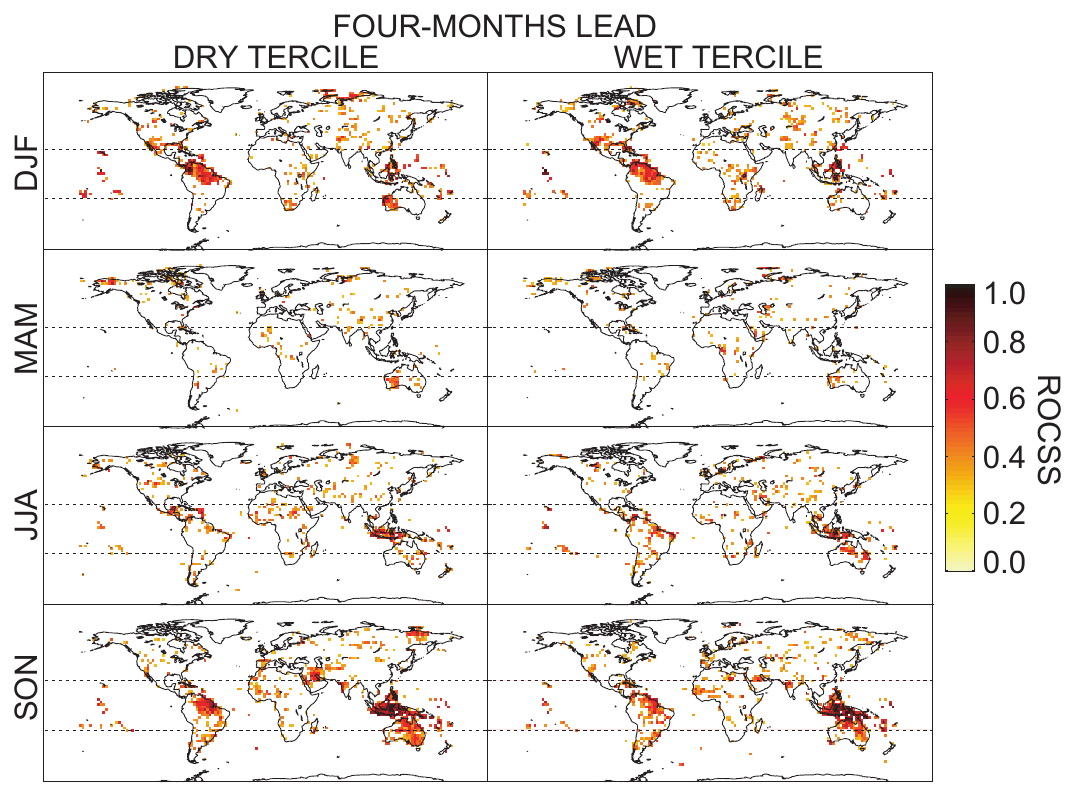}
\caption{As Fig. \ref{f.skill_mm_lt1}, but for the four-months lead predictions.}
\label{f.skill_mm_lt4}\end{center}
\end{figure*}

In order to further analyze the above results in the different regions of the world, global spatial maps of ROCSS were obtained for all the models and the MM. Given its better performance, only results for the MM are reported in the following, for the sake of conciseness. Figs. \ref{f.skill_mm_lt1} and \ref{f.skill_mm_lt4} show the significant skill for the dry (left column) and wet (right column) terciles ---almost no predictability is obtained
for normal conditions--- at one and four months lead-time, respectively, by seasons (in rows).  As can be seen, significant skill is mainly located over the tropics ---being SON (MAM) the most (less) skillful season overall--- and does not decrease noticeably from one to four months lead-time except in MAM (in agreement with Fig. \ref{f.skill_area}).

By seasons, the main skillful regions at one month lead-time in DJF are the gulf of California, south America, central and southern Africa, western Australia and the Pacific islands of Oceania in Melanesia, Micronesia and Polynesia. Except in Africa, where the predictability signal weakens, most of this skill remains at four months lead-time.
In MAM, skill at one month lead-time is located over parts of western U.S.A., northeastern Brazil, parts of the Arabic peninsula, Indochina and the Malay archipelago. However, most of this predictability vanishes at four months lead-time. 
In JJA, central America, northern Brazil, the gulf of Guinea, the Malay archipelago, eastern Australia and the Pacific islands of Oceania are the main skillful regions at one month lead-time. This skill is only  maintained in the Malay archipelago and the Pacific islands of Oceania at four months lead-time.
Finally, one-month lead skill in SON is located over northern south America, a belt in central Africa (especially in the Somali peninsula), parts of Middle East, the Malay archipelago, Australia and the Pacific islands of Oceania.
Moreover, this skill remains unaltered at four months lead-time for all the aforementioned regions except the Somali peninsula, indicating thus a strong predictability signal. 

In the light of the previous results, northern south America and the Malay archipelago seem to be the most skillful regions of the world for seasonal forecasting of precipitation. Note that seasonal predictability in these regions has been analyzed in previous studies \citep{Aldrian2007,Haylock2001}, considering also its derived socio-economic impacts \citep{Kirono1999}.
 
Finally, it is noteworthy to mention that there is a symmetry in the ROCSS found for both dry and wet terciles, leading to similar maps in both cases. The reason for this is that large dry (wet) HIR values partially contribute to low wet (dry) FAR values, since those dry (wet) events correctly forecast are also non-wet (non-dry) events correctly forecast. However, the reverse is not generally true; e.g., non-wet correct forecasts do not necessarily imply a correct forecast of the corresponding dry (or normal) event. From a practical point of view, the positive prediction of a particular event (dry or wet) is more valuable than its negative one (normal-wet or normal-dry, respectively). Thus, besides obtaining the ROCSS maps it is also important to understand the factors driving this skill and how they influence both dry and wet events.  
As we will see in the following section, this is regulated to a great extent by the El Ni\~no phenomenon.

\subsection{El Ni\~no-Driven Skill}\label{s.elnino_skill}
\begin{figure*}[htb]\begin{center}
\noindent\includegraphics[width=0.95 \linewidth]{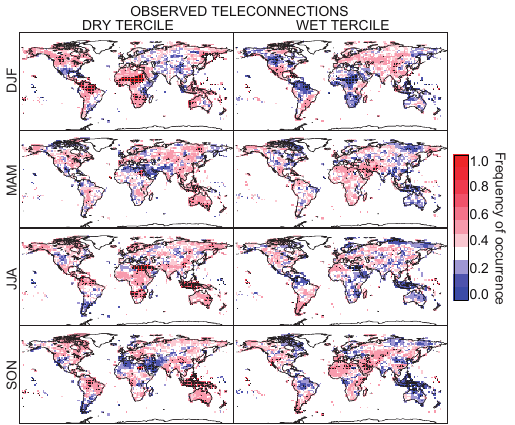}
\caption{Relative frequency of occurrence for the dry and wet terciles (left and right columns, respectively) in El Ni\~no years for the different seasons (in rows). Red (blue) colors correspond to values above (under) $1/3$, the expected  climatological frequency. Black dots indicate significant ---at a 5\% level--- teleconnections, according to a one-sample Z-test for proportions.}
\label{f.telcon_obs}\end{center}
\end{figure*}

Despite the important achievements reached in seasonal forecasting in the last ten years, significant levels of skill for precipitation are only generally found over regions strongly connected with ENSO \citep[see, e.g.,][]{Coelho2006,Barnston2010,Arribas2011,Lim2011,Kim2012a,Kim2012b,Landman2012}, which is known to be the dominant mode of seasonal variability \citep{DoblasReyes2010}. The aim of this section is to analyze to which extent El Ni\~no contributes to the overall (1961-2000) skill found in Sec. \ref{s.global_skill}. To this, we computed the observed El Ni\~no teleconnections in order to look for a possible link between them and the seasonal predictability exhibited by the models ---and consequently, the MM--- (see Figs. \ref{f.skill_mm_lt1} and \ref{f.skill_mm_lt4}).
Teleconnections were also calculated following a tercile-based approach, in terms of the frequencies of occurrence of each category in El Ni\~no periods ---as compared with the climatological frequency $1/3$---. The eleven El Ni\~{n}o years considered were the ones defined in \citet{Pozo-Vazquez2005} as \textit{El Ni\~{n}o Winter} years: 1964, 1966, 1970, 1973, 1977, 1983, 1987, 1988, 1992, 1995 and 1998. Alternatively, we also considered \textit{El Ni\~no Autumn }years ---defined as well  \citet{Pozo-Vazquez2005}---, obtaining similar conclusions (not shown). A one-sample Z-test for proportions was applied to detect those frequencies significantly higher (lower) than $1/3$, which are considered as significant positive (negative) El Ni\~no teleconnections. Fig. \ref{f.telcon_obs} shows the results obtained for the dry and wet terciles (left and right columns, respectively) in the different seasons (in rows). Red/blue for the dry (wet) tercile indicates those grid points where dry/normal-wet (wet/normal-dry) conditions are mainly observed in El Ni\~no years. Overall, our results are in agreement with previous studies \citep[see, e.g.,][]{Ropelewski1987,VanOldenborgh2000,Kayano2009,Shaman2011,Zhang2012,Yadav2013,Zhang2013}.
  
However, rather than analyzing the teleconnections in the different regions of the world, here we focus on the relationship between these teleconnection maps and those reporting the skill (Figs. \ref{f.skill_mm_lt1} and \ref{f.skill_mm_lt4}). Visual inspection reveals that, in general, the skillful areas correspond to those significantly teleconnected with El Ni\~{n}o (black dots in Fig. \ref{f.telcon_obs}), as occurs, for instance, in northern south America in DJF and SON and in Middle East and the Malay archipelago in SON. Moreover, there are no spatially consistent zones of skill which are not teleconnected with this phenomenon.
This suggests that both the seasonal and spatial distribution of the overall skill can be accurately explained by the observed El Ni\~{n}o teleconnections, i.e., the skill found might have a strong El Ni\~{n}o-driven component. 

\begin{figure*}[htb]\begin{center}
\noindent\includegraphics[width=0.95 \linewidth]{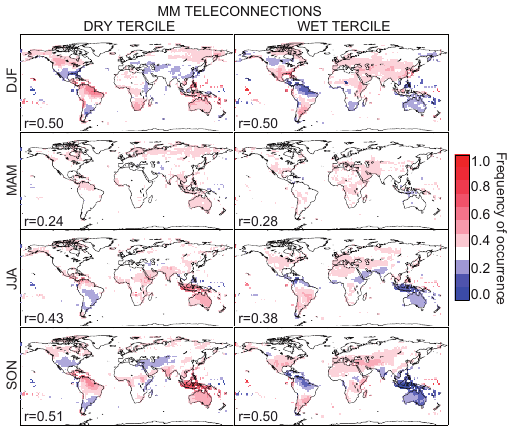}
\caption{El Ni\~{n}o teleconnections reproduced by the MM for the dry and wet terciles (left and right columns, respectively) at one month lead-time. The numbers in each panel show the spatial correlation with the corresponding observed patterns (Fig. \ref{f.telcon_obs}). Note that significance can not be calculated in this case since the independence assumption required by Z-test do not hold due to the interdependence among members within the same model.}
\label{f.telcon_mm}\end{center}
\end{figure*}

Therefore, it is reasonable to think that the models' skill might be determined to a great extent by their ability to properly reproduce the observed El Ni\~{n}o teleconnections. To further analyze this fact, the teleconnections reproduced by the different models and the MM were computed and compared against the observed ones. Note that, although a similar analysis has been done in \cite{Yang2012}, they did not relate their results to the models' performance, which is the aim here. As the observed ones, the simulated teleconnections were also computed in terms of the frequencies of occurrence of the different terciles in El Ni\~{n}o years ---from a total of 11 (El Ni\~no years) $\times$ 9 (members) = 99 values for each single model and  11 (El Ni\~no years) $\times$ 45 (members) = 495 values for the MM---. Note that the independence assumption required by the Z-test does not hold in this case (members are not independent) and, thus, the significance of the results is not computed. Fig. \ref{f.telcon_mm} shows the resulting MM teleconnections for the one-month lead predictions (results are similar for the four months case). 
As can be seen, the predicted patterns are smoother than the observed ones (this feature is less pronounced for the individual models) and are in good agreement with them (see Fig. \ref{f.telcon_obs}) ---numbers in each map show the spatial correlation between the observed and predicted patterns---. Furthermore, agreement is noticeable ---correlations around 0.5--- in the most skillful seasons (SON and DJF) whereas it is poor in the season with the lowest skill (MAM).
These results confirms that the MM appropriately reproduce the observed teleconnections in the different skillful regions and, thus, that the overall skill found in Sec. \ref{s.global_skill} might be determined to a great extent by this ability.

\begin{figure*}[htb]\begin{center}
\noindent\includegraphics[width=0.9 \linewidth]{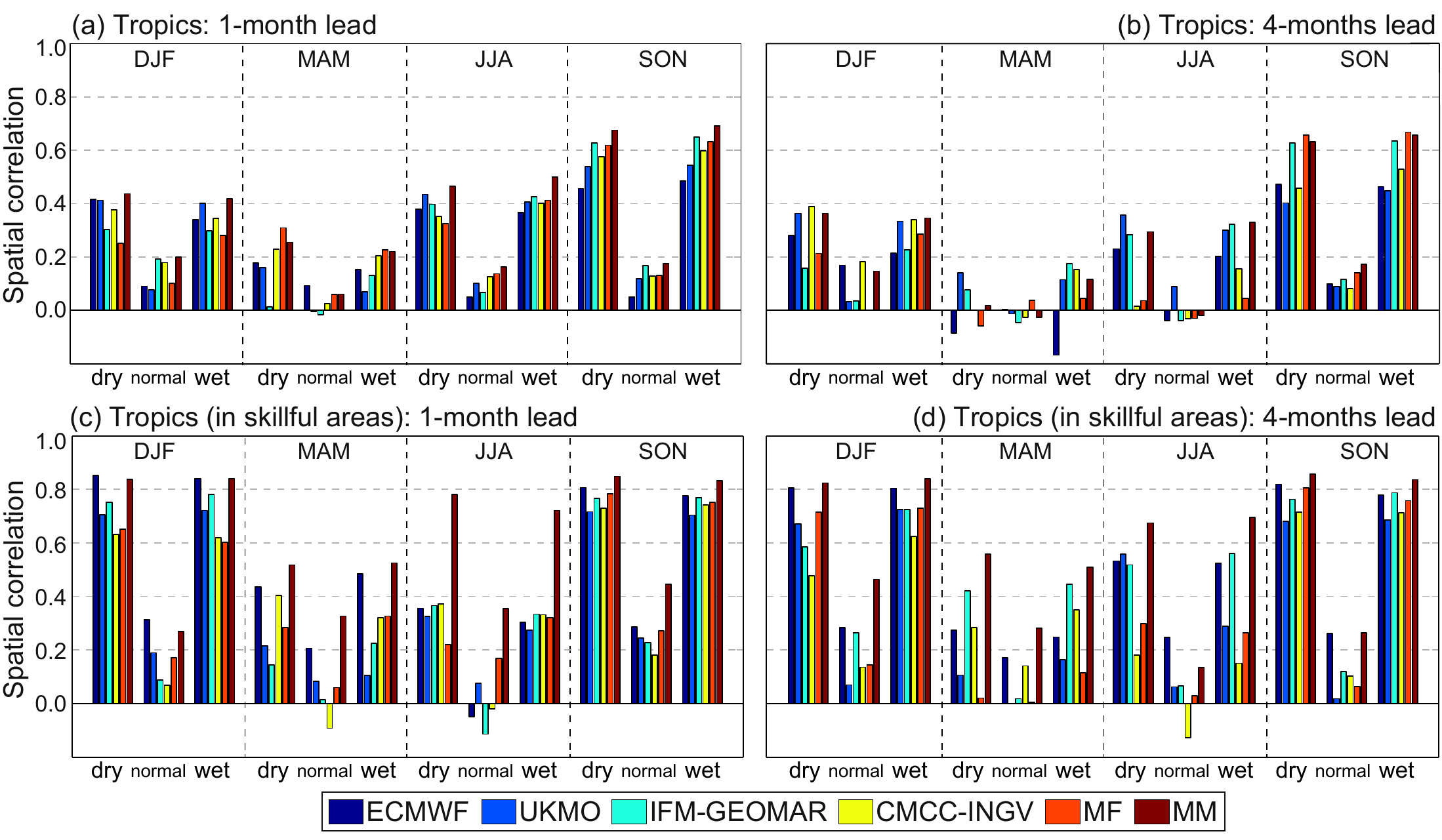}
\caption{Spatial correlation between the observed and predicted El Ni\~{n}o teleconnection patterns in the tropics for (a,c) one- and (b,c) four-months lead seasonal predictions for the five models and the MM (see legend for colors). The panels in the top correspond to the entire tropics, whereas panels in the bottom correspond to those tropical grid points showing significant skill (see Figs. \ref{f.skill_mm_lt1} and \ref{f.skill_mm_lt4}).}
\label{f.spatial_correlation}\end{center}
\end{figure*}

In order to quantify this claim for the rest of models, Fig. \ref{f.spatial_correlation} shows the spatial correlation between the observed and predicted ---for one- and four-months lead forecasts (left and right panels, respectively)--- teleconnections over (a,b) the entire tropics and (c,d) those tropical grid points exhibiting significant skill for the five single models and the MM (see colors in legend).
Results are consistent with the above argumentation, finding the highest (lowest) correlations in the most (less) skillful seasons, DJF and SON (MAM). Moreover, correlations are systematically higher when restricting the analysis to skillful grid points ---e.g., values in SON increase from 0.6 to 0.8---.
In addition, note that this figure is in very much agreement with the results obtained in terms of skillful areas (see Fig. \ref{f.skill_area}a-b).

Finally, we want to mention that we also tried to quantify the contribution of the El Ni\~no phenomenon to the skill by computing the ROCSS in the eleven El Ni\~no years. However, the results turned out to be misleading due to the existence of unobserved tercile categories (e.g., dry events) in some grid points in El Ni\~{n}o periods. In these cases, the ROCSS can be only calculated for the normal and wet events and, thus, do not reflect the ability of the models to predict dry events. For instance, the skill obtained in Middle East in SON completely vanishes when conditioning the analysis to El Ni\~no years (not shown), suggesting an alternative source of skill rather than El Ni\~{n}o for this region. However, the above results clearly show that the skill there is clearly driven by El Ni\~no. Therefore, we want to warn on the use of the ROCSS in cases with small samples where some categories could remain unobserved.

\section{Conclusions}\label{s.conclusions}
The skill of seasonal precipitation forecasts has been assessed worldwide for the forty-years period 1961-2000. To this aim, the ENSEMBLES multi-model seasonal hindcast was considered.  A probabilistic tercile-based validation scheme was applied, obtaining the skill ---in terms of the ROC Skill Score--- and its statistical significance grid point by grid point. Although predictability varies with region, season and lead-time, results indicate that skill is mainly located in the tropics ---with $20$ to $40$\% of significant skillful areas--- rather than in the extra-tropics ---only around the $10$\%---. Overall, DJF and SON (MAM) are the most (less) skillful seasons. In particular, seasonal predictability is especially strong in SON, where the skill found over northern south America, a belt in central Africa, parts of Middle East, the Malay archipelago, Australia and the Pacific islands of Oceania at one month lead-time remains almost unaltered at four-months lead.

The extent to which El Ni\~{n}o contributes to the latter skill was evaluated by means of an analysis of teleconnections. Results reveal that skillful areas are essentially those significantly teleconnected with El Ni\~{n}o, what suggest that the skill found might be driven by this phenomenon. To further analyze this, the performance of the different models to reproduce the El Ni\~{n}o teleconnections was assessed by computing the spatial correlation between the observed and predicted teleconnection patterns. The highest correlations (around 0.6 in the tropics) are found in SON, whereas the lowest (nearly negligible) are found in MAM ---the most and less skillful seasons, respectively---. Moreover, the latter correlations are systematically higher (over 0.8) when restricting the analysis to the skillful areas. These results indicate that the seasonal and spatial distribution of the skill can be explained to a great extent by the model's ability to properly reproduce the observed El Ni\~{n}o teleconnections. 

%
\begin{acknowledgments}
This study was supported by the EU projects QWeCI and SPECS, funded by the European Commission Seventh Framework Research Programme under grant agreements 243964 and  308378, respectively.
The authors want also to acknowledge the ENSEMBLES project, funded by the European Commission Sixth Framework Research Programme through contract GOCE-CT-2003-505539, for the data models' data, which were retrieved from the Meteorological Archival and Retrieval System (MARS) of the ECMWF, \url{http://www.ecmwf.int/services/archive/}. 
\end{acknowledgments}


\end{article}

\end{document}